\documentclass[journal,twocolumn]{IEEEtran}

\usepackage[T1]{fontenc}
\usepackage[latin9]{inputenc}
\usepackage{verbatim}
\usepackage{amsthm}
\usepackage{amsmath}
\usepackage{amssymb}
\usepackage{graphicx}
\usepackage[english]{babel}
\usepackage{subcaption}
\usepackage{authblk}

\makeatletter

\providecommand{\tabularnewline}{\\}

\numberwithin{equation}{section}
\numberwithin{figure}{section}
\numberwithin{table}{section}
\theoremstyle{plain}
\newtheorem{thm}{\protect\theoremname}[section]
  \theoremstyle{definition}
  \newtheorem{defn}[thm]{\protect\definitionname}
  \theoremstyle{remark}
  \newtheorem{rem}[thm]{\protect\remarkname}
  \theoremstyle{plain}
  \newtheorem{cor}[thm]{\protect\corollaryname}
  \theoremstyle{plain}
  \newtheorem{lem}[thm]{\protect\lemmaname}

\makeatother

  \providecommand{\corollaryname}{Corollary}
  \providecommand{\definitionname}{Definition}
  \providecommand{\lemmaname}{Lemma}
  \providecommand{\remarkname}{Remark}
\providecommand{\theoremname}{Theorem}

\begin{document}

\title{Recovery of Sparse Positive Signals on the Sphere from Low Resolution
Measurements}

\author[1]{Tamir Bendory }
\author[1]{Yonina C. Eldar \thanks{This work was funded by the European Union's Horizon 2020 research
and innovation programme under grant agreement ERC-BNYQ, by the Israel Science Foundation under Grant no. 335/14, and by ICore: the Israeli Excellence Center 'Circle of Light'.}}
\affil[1]{Department of Electrical Engineering, Technion -- Israel Institute of Technology, Haifa, Israel. }
\maketitle
\begin{abstract}
This letter considers the problem of recovering a positive stream
of Diracs on a sphere from its projection onto the space of low-degree
spherical harmonics, namely, from its low-resolution version. We suggest
recovering the Diracs via a tractable convex optimization problem.
The resulting recovery error is proportional to the noise level and
depends on the density of the Diracs. We validate the theory by numerical
experiments. 
\end{abstract}

\section{Introduction}

Many applications in engineering and physics consider signals that
lie on spheres (see for instance, \cite{audet2011directional,komatsu2011seven,meyer2001beamforming,jarrett20103d}).
In this letter we consider the problem of recovering a positive stream
of Diracs on the sphere from its low-resolution measurement. The
natural way to model a low-resolution version of a signal on a sphere
is by its projection onto the space of low-degree spherical harmonics,
as will be explained in Section \ref{sec:Mathematical-Model}. 

This work is motivated by the problem of estimating the orientations
of the white matter fibers in the brain using diffusion weighted magnetic
resonance imaging (MRI) \cite{Lazar2015}. It is common to model the
measured signal as a spherical convolution of the underlying distribution
of fiber bundles, called the orientation density function, with the
point spread function of the diffusion tensor imaging sequence which
smears out the fine details of the fibers' distribution. The orientation
density function is modeled as a stream of Diracs on the sphere. The
locations and the positive weights of the Diracs represent the orientations
of the fibers and the partial volume of the fiber within a voxel,
respectively \cite{tournier2004direct,deslauriers2012spherical}.
Therefore, the mathematical model elaborated in Section \ref{sec:Mathematical-Model}
suits this application. 

From the theoretical side, as far as we know, this is the first work
to suggest a stable recovery of positive signals on the sphere
from their low-resolution measurements. This result is part of an
ongoing effort to derive recovery guarantees for super-resolution
of signals in various geometries and settings (see, e.g. \cite{candes2013towards,candes2013super,de2012exact,duval2013exact,moitra2014threshold,demanet2014recoverability,bendory2013Legendre,bendorySOP,bendory2015stable,bendory2013exact}). 

Several papers considered the recovery of 
Diracs on a sphere with general coefficients (not necessarily positive) from their
low-resolution measurements. In \cite{bendory2013exact,bendorySHalgorithm} it was
shown that recovery via convex optimization methods is robust under the
assumption that the Diracs are sufficiently separated (see Theorem
\ref{thm:general}). 
The papers \cite{deslauriers2013sampling,Dokmanic2015} employ a finite
rate of innovation framework to super-resolve Diracs on the sphere.
This approach does not need any assumption on the Diracs' distribution. However, these works have no robustness guarantees.
Additionally, in \cite{bendorySHalgorithm} it was proven that
separation is necessary for robust recovery in the
presence of noise \emph{by any method}.

Following \cite{morgenshtern2014stable}, we
show that if the Diracs are known to be positive then the separation
condition can be replaced by a weaker condition called Rayleigh regularity,
 which quantifies the density of the Diracs. We suggest
recovering the Diracs via a tractable convex optimization problem.
The resulting recovery error is proportional to the noise level and depends
on the Rayleigh regularity of the signal.

The letter is organized as follows. In Section \ref{sec:Mathematical-Model} we formulate the problem and present necessary mathematical background. Section \ref{sec:Main-result} presents our main result, which is proved in Section \ref{sec:Proof-of-Theorem}.  Section \ref{sec:Numerical-experiments} shows some numerical experiments, which corroborate the theoretical results, and Section \ref{sec:conclusion} concludes the letter.

\section{Problem Formulation and Background \label{sec:Mathematical-Model}}

Spherical harmonics play a key role in the analysis of signals in
a vast variety of tasks, analysis methods and sampling theorems, see
for instance; \cite{arridge1999optical,rafaely2005analysis,cohen2006quantum,mcewen2011novel,leistedt2012exact,iglewska2014continuous}.
Let $\mathcal{Y}_{n}\left(\mathbb{S}^{2}\right)$ denote the space
of homogeneous spherical harmonics of degree $n$, which is the restriction to the bivariate unit sphere of the homogeneous harmonic polynomials of degree $n$ in $\mathbb{R}^3$. 

Any point on the bivariate unit sphere $\mathbb{S}^{2}$ is parametrized
by $x:=\left(\phi,\theta\right)\in[0,2\pi)\times\left[0,\pi\right]$.
The distance between two points $x_{i},x_{j}\in\mathbb{S}^2$ is measured as 
\begin{equation}
\rho\left(x_{i},x_{j}\right):=\arccos\left(x_{i}\cdot x_{j}\right).\label{eq:rho}
\end{equation}
 Let 
\[
Y_{n,k}=A_{n,k}e^{jk\phi}P_{n,k}\left(\cos\theta\right),\quad k=-n,\dots,n,
\]
 be an orthonormal basis of $\mathcal{Y}_{n}\left(\mathbb{S}^{2}\right)$,
where $P_{n,k}\left(x\right)$ is an associated Legendre polynomial
of degree $n$ and order $k$, and 
\begin{equation*}
A_{n,k}:=\sqrt{\frac{2n+1}{4\pi}\frac{\left(n-\left|k\right|\right)!}{\left(n+\left|k\right|\right)!}}.
\end{equation*}
The functions $\left\{ Y_{n,k}\right\} $ are the eigenfunctions of
the Laplacian on $\mathbb{S}^{2}$, and thus can be understood as
the extension of Fourier analysis on the sphere. Any function $g\in L_{2}\left(\mathbb{S}^{2}\right)$
can be expanded as \cite{atkinson2012spherical}
\begin{equation*}
g(x)=\sum_{n=0}^{\infty}\sum_{k=-n}^{n}\left\langle g,Y_{n,k}\right\rangle Y_{n,k}(x).
\end{equation*} 

In this work, we consider a discrete positive signal of the form 
\begin{equation}
f[x]=\sum_{m=1}^{M}c_{m}\delta\left[x-x_{m}\right],\quad   c_{m}>0,\label{eq:signal}
\end{equation}
where $\delta\left[x\right]$ is the Kronecker delta function and $X:=\left\{ x_{m}\right\} $ is the signal's support. We
assume that the signal lies on some predefined grid $\mathbb{S}_{L}^{2}\subset\mathbb{S}^{2}$
and that any pair of points on the grid $x_{i},x_{j}\in\mathbb{S}_{L}^{2}$
satisfy $\rho\left(x_{i},x_{j}\right)\geq1/L$ for some $L\geq1/\pi$.
The higher $L$ is, the larger the target resolution we want to achieve.

The information we have on the signal is its projection onto the space
of the low $N$ spherical harmonics 
\begin{equation}
\begin{split}
&y_{n,k}=\left\langle f,Y_{n,k}\right\rangle +\eta_{n,k}, \\ &n=0,\dots,N,\quad  k=-n,\dots, n,\label{eq:meas}
\end{split}
\end{equation}
where $\eta:=\left\{ \eta_{n,k}\right\} $ is some noise or model
mismatch which is assumed to be bounded. In matrix notation we may
write 
\begin{equation}
y=F_{N}f+\eta\iff s:=F_{N}^{*}y=P_{N}f+F_{N}^{*}\eta,\label{eq:meas_matrix}
\end{equation}
where $y:=\left\{ y_{n,k}\right\} $, $F_{N}$ is a linear operator
mapping a signal to its low $N$ spherical harmonic coefficients,
and the adjoint operator is given by $F_{N}^{*}z(x)=\sum_{n\leq N,\left|k\right|\leq n}z_{n,k}Y_{n,k}(x).$
The operator $P_{N}=F_{N}^{*}F_{N}$ is the orthogonal projection
onto the space of spherical harmonics of degree $N$, denoted by $V_{N}$.
We aim to recover the sets $\left\{ c_{m}\right\} ,\thinspace\left\{ x_{m}\right\} $
from the noisy low-resolution measurements (\ref{eq:meas}).

In recent papers \cite{bendory2013exact,bendorySHalgorithm}, it was
shown that signals of the form (\ref{eq:signal}) with general coefficients
(namely, not necessarily positive values) can be recovered robustly
from $V_{N}$ by solving a tractable convex program. This holds provided
that the signal's support $X$ satisfies the following separation
condition:
\begin{defn}
\label{def:separation}A set of points $X\subset\mathbb{S}^{2}$ is
said to satisfy the minimal separation condition if

\[
\min_{x_{i},x_{j}\in X,\thinspace i\neq j}\rho\left(x_{i},x_{j}\right)\geq\frac{\nu}{N},
\]
for a fixed \emph{separation constant} $\nu$ that does not depend
on $N$, where $\rho$ is defined in (\ref{eq:rho}). 
\end{defn}
We will also make use of the notion of the super-resolution factor (SRF).
The SRF quantifies the ratio between the resolution we want to achieve,
specified by the grid spacing $1/L$, and the resolution we measure,
namely, 
\begin{equation}
SRF:=\frac{L}{N}.\label{eq:srf}
\end{equation}
The main result of \cite{bendorySHalgorithm} then states the following:
\begin{thm}
\label{thm:general}Let $X=\left\{ x_{m}\right\} \subset\mathbb{S}_{L}^{2}$
be the support of a signal of the form (\ref{eq:signal}) with general
coefficients $c_{m}\in\mathbb{R}$. Let $\left\{ y_{n,k}\right\} $
be as in (\ref{eq:meas}) with $\left\Vert \eta\right\Vert _{\ell_{2}}\leq\delta$.
For sufficiently large $L$, if $X$ satisfies the separation condition
of Definition \ref{def:separation}, then the solution $\hat{f}$
of 
\begin{equation}
\min_{g\in\mathbb{S}_{L}^{2}}\left\Vert g\right\Vert _{\ell_{1}}\mbox{ subject to }\left\Vert y-F_{N}g\right\Vert _{\ell_{2}}\leq\delta,\label{eq:ell1}
\end{equation}
satisfies 
\[
\left\Vert \hat{f}-f\right\Vert _{\ell_{1}}\le C_{0}SRF^{2}\delta,
\]
for some fixed constant $C_{0}$.\end{thm}
\begin{rem}
The minimal separation constant $\nu$ was evaluated numerically to
be $2\pi$. The separation of $\frac{2\pi}{N}$ coincides with the
spatial resolution of signals on spheres \cite{rafaely2004plane}. 
\end{rem}
The aim of this letter is to derive a stability result for the recovery
of positive signals on the sphere from their low-resolution measurements.
In this case, as presented in the next section, the separation condition
can be replaced by a weaker condition called Rayleigh regularity.

\section{\label{sec:Main-result}Main result}

In \cite{bendory2013exact}, it was proven that a positive signal
on the sphere with $M\leq N/2$ (i.e. maximal cardinality of $N/2$)
can be perfectly recovered from its noiseless projection onto $V_{N}$ by solving a convex optimization problem. A general signal of cardinality $M$ can be recovered from $V_{M+1+\sqrt{M+1}}$ by an algebraic approach \cite{Dokmanic2015}.
However, both recovery results are not stable in the presence of noise. 

To derive a stability result for the positive case, we use the notion of Rayleigh regularity.
A univariate signal with Rayleigh regularity $r$ has at most $r$
spikes within a resolution cell of size $\frac{\mu}{N}$ for a separation
constant $\mu$. In the multidimensional case, the definition of Rayleigh
regularity is less intuitive (see discussion in \cite{bendoryPositiveSOP})
and can be interpreted as a density measure of the signal. For $r=1$,
the definition coincides with Definition \ref{def:separation}. 
\begin{defn}
\label{def:2Dray} We say that the set $\mathcal{P}\subset\mathbb{S}_{L}^{2}$
is Rayleigh-regular with parameters $(\mu,r\thinspace;\thinspace N,L)$
and write $\mathcal{P}\in\mathcal{R}^{idx}(\mu,r\thinspace;\thinspace N,L)$
if 
\begin{itemize}
\item $\mathcal{P}\mathcal{=}\mathcal{P}_{1}\cup\dots\cup\mathcal{P}_{r}$
and $\mathcal{P}_{i}\cap\mathcal{P}_{j}=\emptyset$ for all $i\neq j$,
\item for all $i=1,\dots,r,$ $\mathcal{P}_{i}$ satisfies the separation
condition of Definition \ref{def:separation} with constant
$\mu$.
\end{itemize}
We denote the set of positive signals of the form (\ref{eq:signal})
with support $X\in\mathcal{R}^{idx}(\mu,r\thinspace;\thinspace N,L)$
as $\mathcal{R}_{+}(\mu,r\thinspace;\thinspace N,L)$.
\end{defn}
The notion of Rayleigh regularity was used in \cite{morgenshtern2014stable}
to derive stability results for the recovery of positive signals
from their low-degree Fourier coefficients. Our results can be seen
as an extension to signals on spheres. 

In the sequel, we assume that the noise satisfies $\left\Vert F_{N}^{*}\eta\right\Vert _{\ell_{1}}\leq\delta$
and suggest to recover the signal by solving the feasibility (convex)
problem 
\begin{equation}
\mbox{find }g\in\mathbb{S}_{L}^{2}\quad\mbox{subject to}\quad\left\Vert s-P_{N}g\right\Vert _{\ell_{1}}\leq\delta,\thinspace g\geq0.\label{eq:cvx}
\end{equation}
Now, we are ready to present our main theorem. The theorem states
that by solving the convex program (\ref{eq:cvx}), one can stably
recover a positive signal on the sphere from its low-resolution
measurements. The recovery error is proportional to the noise level
and depends on the Rayleigh regularity of the signal. 
\begin{thm}
\label{thm:main}Let $f\in\mathcal{R}_{+}(\nu r,r\thinspace;\thinspace N,L)$
be of the form (\ref{eq:signal}) and consider the measurement model
(\ref{eq:meas}). Then, for sufficiently large SRF, any solution $\hat{f}$
of (\ref{eq:cvx}) satisfies 
\[
\left\Vert \hat{f}-f\right\Vert _{\ell_{1}}\leq4C_{1}^{-r}r^{2r}SRF^{2r}\delta,
\]
for some fixed constant $C_{1}>0$. \end{thm}
\begin{cor}
In the noiseless case, $\delta=0$, the recovery is exact. 
\end{cor}
As we show in the simulations, minimizing the $\ell_{1}$ norm in
(\ref{eq:cvx}) among all feasible solutions results in a low recovery
error.

\section{\label{sec:Proof-of-Theorem}Proof of Theorem \ref{thm:main}}

The proof exploits the technique presented in \cite{morgenshtern2014stable}
(see also \cite{bendoryPositiveSOP}). We commence by presenting the
following Lemma which is a direct consequence of the construction
in \cite{bendory2013exact}:
\begin{lem}
\label{lem:q_general} Suppose that the set $\Xi=\left\{ \xi_{m}\right\} \subset\mathbb{S}^{2}$
satisfies the separation condition of Definition \ref{def:separation}.
Then, for sufficiently large $N$ there exists a polynomial $q\in V_{N}$
obeying $q(\xi)\leq1$ and 
\begin{eqnarray*}
q\left(\xi_{m}\right) & = & 0,\quad\forall\xi_{m}\in\Xi,\\
q\left(\xi\right) & \geq & C_{1}N^{2}\rho\left(\xi,\xi_{m}\right)^{2},\quad\rho\left(\xi,\xi_{m}\right)\leq\sigma/N,\thinspace\xi_{m}\in\Xi,\\
q\left(\xi\right) & \geq & C_{2},\quad\mbox{if }\rho\left(\xi,\xi_{m}\right)>\sigma/N\mbox{,\thinspace\ensuremath{\forall}\ensuremath{\ensuremath{\xi_{m}\in\Xi}}},
\end{eqnarray*}
for constants $\sigma,C_{1}>0$ and $0<C_{2}<1$.
\end{lem}
Set $h:=\hat{f}-f\subset\mathbb{S}_{L}^{2}$ where $\hat{f}$ is the
solution of the convex program (\ref{eq:cvx}). Let $\mathcal{H}:=\left\{ x\in\mathbb{S}_{L}^{2}\thinspace:\thinspace h\left[x\right]<0\right\} $ and thus $\mathcal{H}\subseteq X$.
By assumption $f\in\mathcal{R}_{+}(\nu r,r\thinspace;\thinspace N,L)$
 and consequently $\mathcal{H}\in\mathcal{R}^{idx}(\nu r,r\thinspace;\thinspace N,L)$.
Therefore, by Definition \ref{def:2Dray}, the set $\mathcal{H}$
can be presented as a disjoint union of $r$ sets $\mathcal{H}=\cup_{i=1}^{r}\mathcal{H}_{i},$
where $\mathcal{H}_{i}\in\mathcal{R}^{idx}(\nu r,1\thinspace;\thinspace N,L)$
for all $i=1,\dots,r$. Observe that by simple rescaling%
\footnote{We assume here that $N/r$ is an integer for clarity. This assumption
is not necessary for the results to hold.%
} $\tilde{N}=N/r$ we have 
\begin{equation*}
\mathcal{R}^{idx}(\nu r,1\thinspace;\thinspace N,L)=\mathcal{R}^{idx}(\nu ,1\thinspace;\thinspace N/r,L).
\end{equation*}
Then, for each set $\mathcal{H}_{i}$ there exists
an associated interpolating polynomial $q_{i}(x)\in V_{N/r}$ as given
in Lemma \ref{lem:q_general}.

The key ingredient of the proof is the following construction: 
\[
\grave{q}\left(x\right)=\prod_{i=1}^{r}q_{i}\left(x\right)-\alpha,
\]
where $\alpha>0$ is a constant to be determined later. A product
of spherical harmonics of degrees $N_{1},N_{2}$ is a spherical harmonic
of degree $N_{1}+N_{2}$ and the computation of the corresponding
representation is known as Clebsch-Gordan. Therefore, $\grave{q}\in V_{N}.$
We denote by $\grave{q}_{L}[x]$ the restriction of $\grave{q}(x)$
to the grid $\mathbb{S}_{L}^{2}$.

By construction, for all $x\in\mathcal{H}$ we have $q_{i}\left(x\right)=0$
for some $i=1,\dots,r$. Therefore, 
\[
\grave{q}_{L}\left[x\right]=\prod_{i=1}^{r}q_{i}\left(x\right)-\alpha=-\alpha.
\]
Additionally, for sufficiently large SRF (see (\ref{eq:srf})) we
get that for all $x\in\mathbb{S}_{L}^{2}\backslash\mathcal{H}$, 
\begin{eqnarray*}
\grave{q}_{L}\left[x\right] & \geq & C_{1}^{r}L^{-2r}\left(\frac{N}{r}\right)^{2r}-\alpha\\
 & = & C_{1}^{r}r^{-2r}SRF^{-2r}-\alpha.
\end{eqnarray*}
By setting 
\begin{equation}
\alpha:=\frac{1}{2}C_{1}^{r}r^{-2r}SRF^{-2r}<1/2,\label{eq:alpha}
\end{equation}
we conclude that 
\begin{eqnarray}
\grave{q}_{L}\left[x\right] & = & -\alpha\quad\forall x\in\mathcal{H},\label{eq:q_alpha}\\
\grave{q}_{L}\left[x\right] & \geq & \alpha,\quad\forall x\in\mathbb{S}_{L}^{2}\backslash\mathcal{H}.\nonumber 
\end{eqnarray}

Once we have constructed the appropriate polynomial $\grave{q}_L$,
the rest of the proof follows directly. On the one hand, by (\ref{eq:meas_matrix})
and (\ref{eq:cvx}) we get 
\begin{eqnarray}
\left|\left\langle \grave{q}_{L},h\right\rangle \right| & =&  \left|\left\langle P_{N}\grave{q}_{L},h\right\rangle \right|\nonumber \\ &=&\left|\left\langle \grave{q}_{L},P_{N}h\right\rangle \right| \nonumber \\ &\leq & \label{eq:lower_bound}
\left\Vert \grave{q}_{L}\right\Vert _{\ell_{\infty}} \left\Vert P_{N}\left(\hat{f}-f\right)\right\Vert _{\ell_{1}}  \\ &\leq &
\left\Vert \grave{q}_{L}\right\Vert _{\ell_{\infty}}\left(\left\Vert P_{N}\hat{f}-s\right\Vert _{\ell_{1}}+\left\Vert s-P_{N}f\right\Vert _{\ell_{1}}\right) \nonumber\\
&  \leq &  2\delta.\nonumber 
\end{eqnarray}
On the other hand, combining (\ref{eq:q_alpha}) with the fact that  $\grave{q}_{L}$ and $h$ have the same sign
pattern on $\mathbb{S}_{L}^{2}$ we have
\begin{eqnarray}
\left|\left\langle \grave{q}_{L},h\right\rangle \right| & = & \left| \sum_{x\in\mathbb{S}_{L}^{2}}\grave{q}_{L}[x]h[x]\right|\nonumber \\ 
&=&\sum_{x\in\mathbb{S}_{L}^{2}}\left|\grave{q}_{L}[x]\right|\left|h[x]\right| \label{eq:upper_bound} \\
 & \geq & \alpha\left\Vert h\right\Vert _{\ell_{1}}.\nonumber
\end{eqnarray}
Combining (\ref{eq:lower_bound}),(\ref{eq:upper_bound}) and (\ref{eq:alpha})
we conclude that 
\begin{eqnarray*}
\left\Vert h\right\Vert _{\ell_{1}} & \leq & \frac{2\delta}{\alpha} \\ &=&4C_{1}^{-r}r^{2r}SRF^{2r}\delta.
\end{eqnarray*}

\section{\label{sec:Numerical-experiments}Numerical experiments}

We now verify the theoretical
results of this paper via numerical experiments. The convex optimization problems were solved
using CVX \cite{cvx}. In all experiments, we set the separation constant
$\nu$ to be $\frac{5\pi}{2}$ and chose a uniform grid 
\begin{equation*}
\mathbb{S}_{L}^{2}:=\left\{ \left(\frac{2\pi q}{L},\frac{\pi p}{L}\right)\thinspace:\thinspace\left(q,p\right)\subset\left[0,\dots,L-1\right]^{2}\right\} .
\end{equation*}

\begin{figure*}

\begin{minipage}[h]{0.7\columnwidth}
\begin{subfigure}[h]{0.7\columnwidth}
       \includegraphics[scale=0.3]{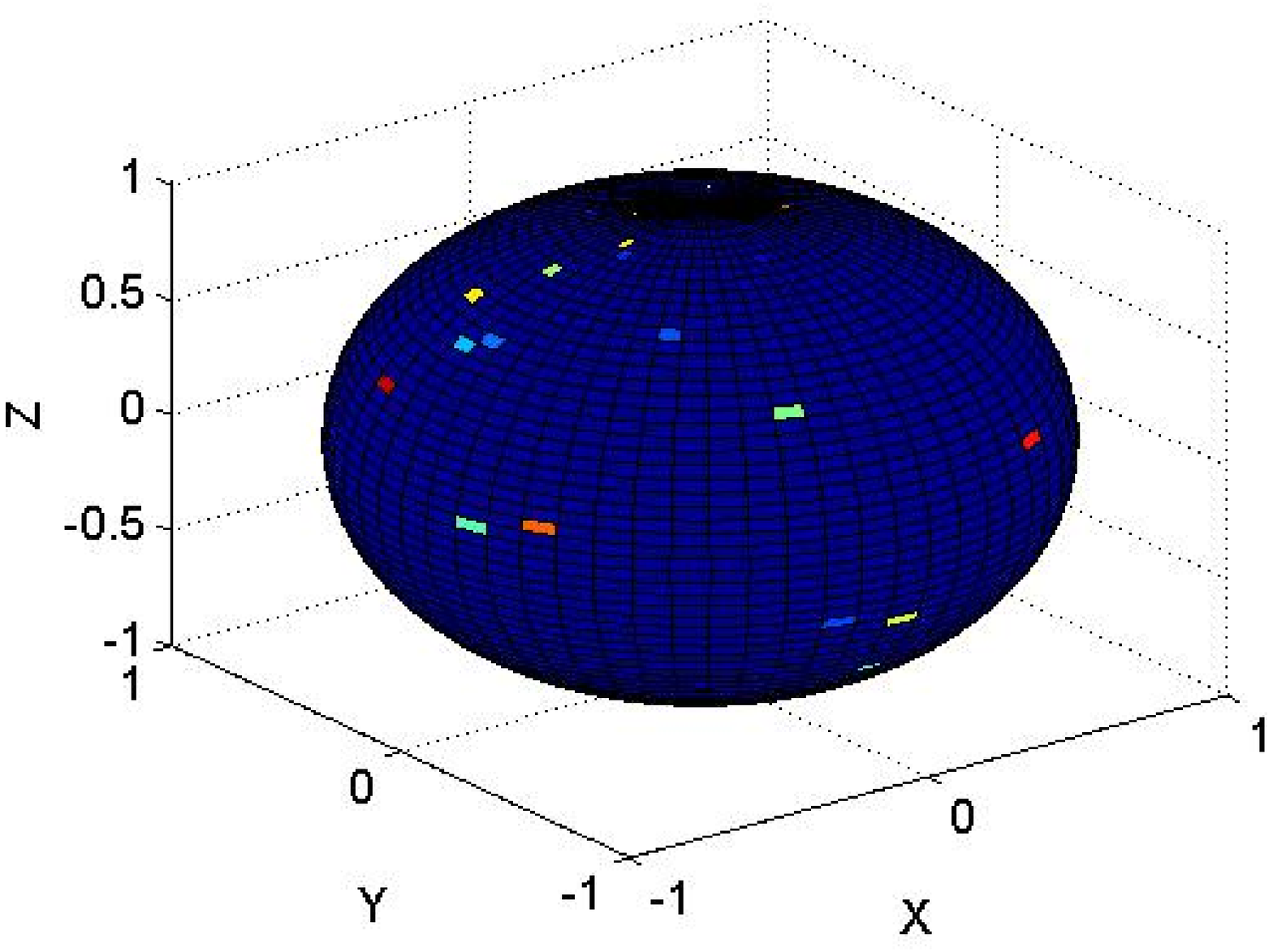}
        \caption{The underlying signal.}
    \end{subfigure}
\end{minipage}
\begin{minipage}[h]{0.7\columnwidth}
\begin{subfigure}[h]{0.7\columnwidth}
\includegraphics[scale=0.3]{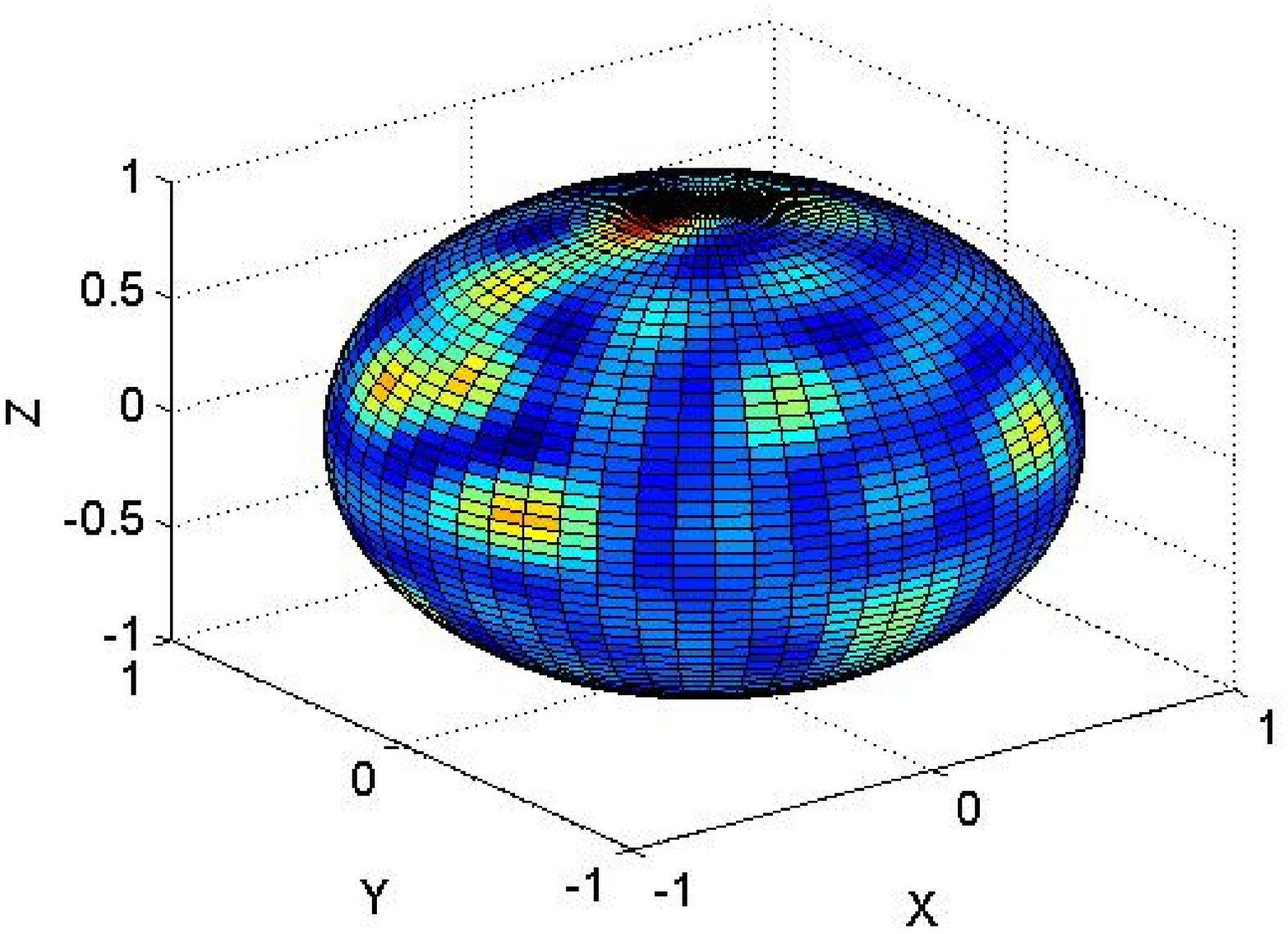}        \caption{Noisy measurements.}
    \end{subfigure}
    \end{minipage}
\begin{minipage}[h]{0.7\columnwidth}
    \begin{subfigure}[h]{0.7\columnwidth}
       \includegraphics[scale=0.3]{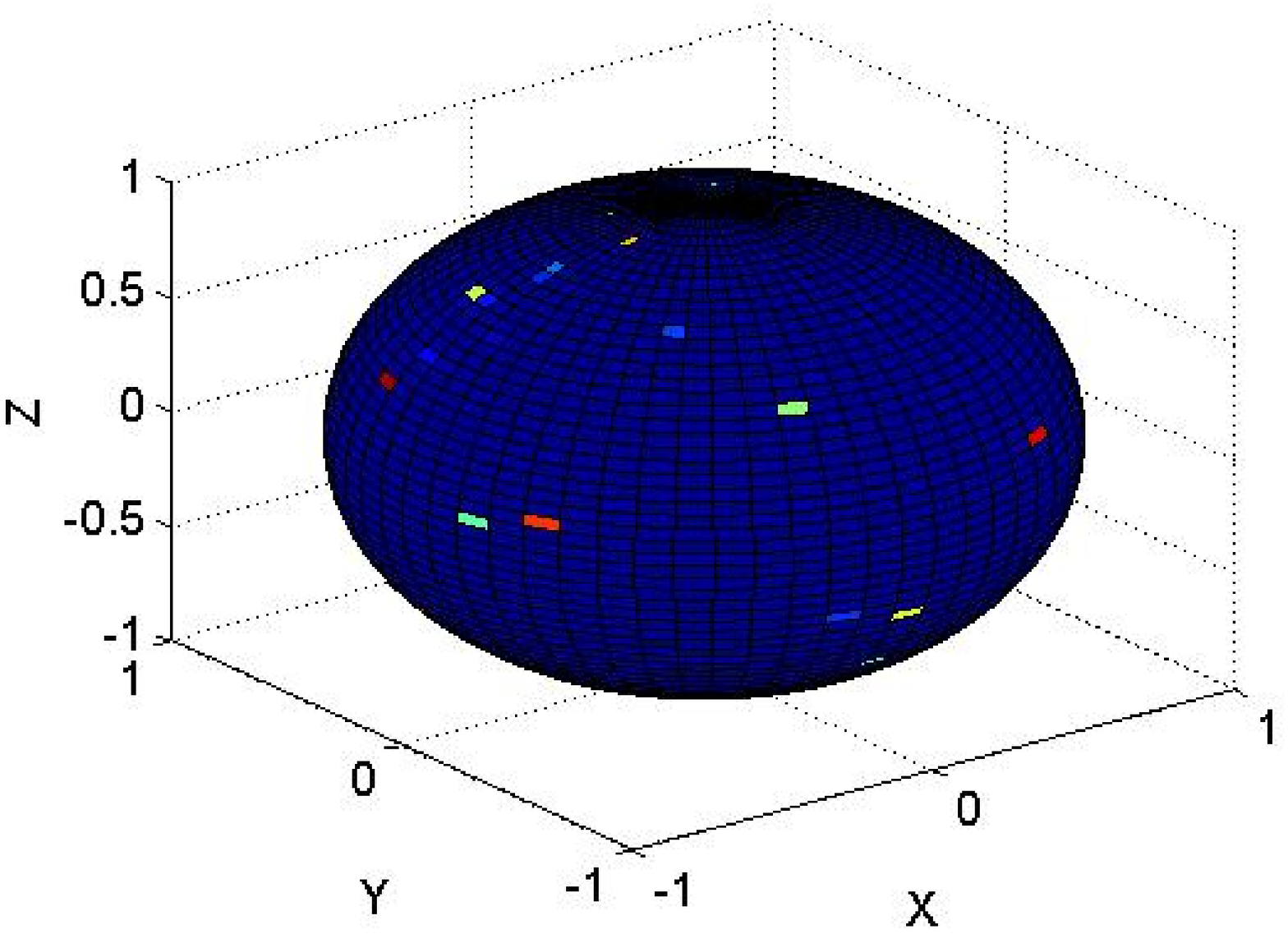}
        \caption{The recovered signal.}
    \end{subfigure}
            \end{minipage}

\protect\caption{\label{fig:recovery_example}An example for the recovery of a signal on the sphere
 from its projection onto $V_N$ with parameters $L=60,\thinspace N=15,$ SRF=4, $r=3, M=41$ and SNR=30 db. }
\end{figure*}

\begin{figure}[h]
\begin{centering}
\includegraphics[scale=0.4]{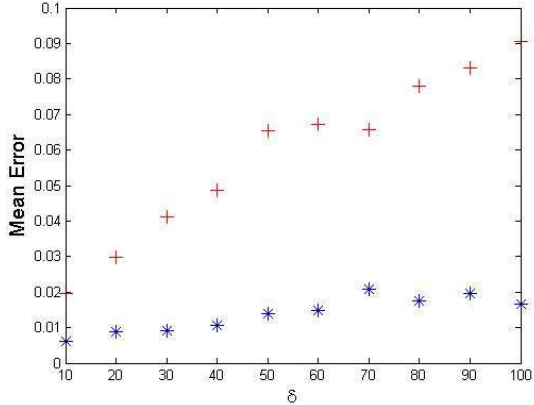}
\par\end{centering}

\protect\caption{\label{fig:errorVSnoise}The mean $\ell_{1}$ recovery error (over
10 experiments) as a function of the noise level with  parameters  $N=12,\thinspace L=50$
and $r=2$. The blue asterisks and the red crosses represent the recovery
error with and without minimizing the $\ell_{1}$ norm among all solutions of (\ref{eq:cvx}), respectively. }
\end{figure}

\begin{table}
\begin{centering}
\begin{tabular}{|c|c|c|c|c|}
\hline 
  Rayleigh regularity  & $r=1$ & $r=2$ & $r=3$ & $r=4$\tabularnewline
\hline 
\hline 
Mean recovery error & 0.0026  & 0.0148  & 0.0285  & 0.0584 \tabularnewline
\hline 
Max recovery error & 0.0059 & 0.0365 & 0.0452 & 0.0699\tabularnewline
\hline 
\end{tabular}
\par\end{centering}

\protect\caption{\label{tab:1}Mean $\ell_{1}$ recovery error (over 10 experiments for each value) 
of the solutions of (\ref{eq:cvx})  with minimal $\ell_{1}$ norm
 as a function of the signal's Rayleigh regularity with the parameters
$L=50,\thinspace N=12,\thinspace SRF\thickapprox4$ and SNR=30 db. }
\end{table}

The signal support was generated as a union of $r$ disjoint sets
that were drawn randomly on the sphere, while keeping the separation
requirements of Definition \ref{def:2Dray}. For each support location,
an associated amplitude was drawn randomly from a uniform distribution
on the interval $(0,10]$. Then, we computed the projection of the
signal onto $V_{N}$ and added an iid normal additive noise. 

We solved the feasibility convex program (\ref{eq:cvx}) and chose the solution
with minimal $\ell_{1}$ norm over all feasible solutions. %
Figure \ref{fig:recovery_example} presents a recovery example of
a signal with Rayleigh regularity of $r=3$ and $M=41$ in a noisy environment
of $SNR=30$ db.
In Figure \ref{fig:errorVSnoise} we compare the output of the CVX
program as a function of the noise level, with and without minimizing the $\ell_{1}$ norm among all
feasible solutions. The recovery error was computed as the normalized
$\ell_{1}$ error, i.e. 
\begin{equation*}
\mbox{error}=\frac{1}{L^{2}}\sum_{p,q=1}^{L}\left|\hat{f}[q,p]-f[q,p]\right|.
\end{equation*}
Table \ref{tab:1} shows the mean recovery error as a function of
the Rayleigh regularity parameter $r$ with $SNR=30$ and $SRF\cong4$.

\section{Conclusion} \label{sec:conclusion}
 
 In this letter, we  proved that a discrete positive stream of Diracs on a sphere can be recovered robustly from its low-resolution measurements by solving a tractable convex optimization problem. The recovery error is proportional to the noise level and depends on the distribution of the Diracs on the sphere.
  
In practice, signals that have sparse representation in a continuous dictionary might not have sparse representation after discretization \cite{chi2011sensitivity}. An obvious technique to alleviate this basis mismatch is by fine discretization, which will increase the recovery error significantly according to Theorem \ref{thm:main}. Proving a  version of Theorem \ref{thm:main} for continuous signals is therefore an important extension for future work.


\end{document}